\documentclass[conference]{IEEEtran}
\usepackage{cite}
\usepackage{amsmath,amssymb,amsfonts}
\usepackage{algorithmic}
\usepackage{graphicx}
\usepackage{textcomp}
\usepackage{multicol}
\usepackage{xcolor}
\usepackage{etoolbox}
\usepackage{caption}
\usepackage[bookmarks=false]{hyperref}
\hypersetup{draft}
\usepackage{makecell}
\usepackage{float}
\def\BibTeX{{\rm B\kern-.05em{\sc i\kern-.025em b}\kern-.08em
    T\kern-.1667em\lower.7ex\hbox{E}\kern-.125emX}}
\begin{document}

\title{AutoCluster, AutoTopicModeling, AutoTrendAnalysis: A Complete AutoML Pipeline for Predicting Emerging Trends\\
\thanks{Identify applicable funding agency here. If none, delete this.}
}

\author{
    \IEEEauthorblockN{
        Ahmed Abolfadl\textsuperscript{1}, 
        Marwa Mahmoud Abla\textsuperscript{2}, 
        Mervat Abu-Elkheir\textsuperscript{1}, 
        Maggie Mashaly\textsuperscript{2}
    }
    \\
    \IEEEauthorblockA{
        \textsuperscript{1}Faculty of Media Engineering and Technology, German University in Cairo, Cairo, Egypt \\
        \textsuperscript{2}Faculty of Information Engineering and Technology, German University in Cairo, Cairo, Egypt
    }
    \\
    \IEEEauthorblockA{
        ahmed.abuelfadel@student.guc.edu.eg, \\
        \{marwa.abla, mervat.abuelkheir, maggie.ezzat\}@guc.edu.eg
    }
}

\maketitle
\begin{abstract}
Predicting emerging trends is vital for businesses, researchers, and policymakers; yet traditional approaches often lack scalability and adaptability. This paper presents a trend prediction framework based on Automated Machine Learning (AutoML), designed to extract insights from textual datasets with temporal attributes. The system ingests subject-specific textual entries accompanied by a date field. The pipeline begins with preprocessing and embedding, followed by AutoClustering, which uses meta-learning to select the optimal clustering algorithm. AutoTopicModeling then applies successive halving to identify the best topic modeling method—Latent Dirichlet Allocation (LDA), Latent Semantic Analysis (LSA), BERTopic, or Non-negative Matrix Factorization (NMF)—based on the coherence score for each cluster. For trend forecasting, AutoTrendAnalysis evaluates multiple models—Facebook Prophet, AutoRegressive Integrated Moving Average (ARIMA), Seasonal-Trend decomposition using Loess (STL), and Long Short-Term Memory (LSTM)—selecting the most accurate based on Root Mean Square Error (RMSE), either through successive halving or exhaustive comparison. Topics are classified as strong signals, weak signals, or noise based on forecasting outcomes, enabling the identification of emerging trends. By automating clustering, topic modeling, and time series forecasting, this research enhances trend prediction accuracy while reducing manual effort. The proposed system offers a scalable and user-friendly solution suitable for real-time applications and stakeholders with limited machine learning expertise. Experimental results demonstrate that the proposed system's best trial achieves a final RMSE of 7.099, indicating high predictive accuracy.
\end{abstract}


\section{Introduction}
The ability to detect and forecast emerging trends has become a strategic imperative across sectors such as finance, technology, public health, and social media. As markets and consumer preferences evolve rapidly, traditional trend analysis methods often fall short in handling the scale and complexity of modern data.

Machine learning offers a powerful solution for extracting patterns from large datasets; however, its widespread adoption is obstructed by the significant technical expertise required. In particular, traditional machine learning approaches to trend prediction~\cite{tarek2024} involve numerous manual decisions by practitioners, such as selecting appropriate models and tuning hyperparameters—tasks that are both time-consuming and demand substantial expertise.

This research proposes the use of AutoML to democratize access to trend prediction tools by automating model selection, hyperparameter tuning and optimization. The goal is to build a fully automated framework capable of identifying, modeling, and forecasting emerging trends from large-scale textual datasets. By minimizing human intervention and maximizing adaptability, such a system can support faster, more accurate, and more scalable trend forecasting.

The proposed pipeline consists of three core components. First, it employs meta-learning to automatically select the best clustering algorithm for unstructured text data. Second, AutoML techniques are used to identify the most coherent topics within each cluster through successive halving. Finally, time series forecasting models assess the temporal evolution of each topic, categorizing them into strong signals, weak signals, or noise.

This end-to-end system is designed for real-time adaptability, continuously updating insights as new data streams in. It holds potential for integration into various domains requiring timely decision-making, such as market analysis, public policy, and innovation forecasting. By reducing technical overhead and enabling broader participation in data-driven decision-making, this work contributes a scalable and accessible solution for trend analysis in an increasingly dynamic world.
\section{Literature Review}

Contributions to AutoML include \cite{he2019automl}, which introduces a framework that automates both model selection and hyperparameter optimization, demonstrating its ability to outperform traditional machine learning models in terms of prediction accuracy and efficiency. Similarly, \cite{domhan2015benchmarking} examines the effectiveness of automated hyperparameter optimization, emphasizing its role in creating well-calibrated models, thereby enhancing the rapid development of machine learning solutions.

Authors of \cite{boutaleb2024bertrend} introduced BERTrend, a neural topic modeling framework designed for detecting emerging trends in large-scale text corpora. BERTrend leverages BERTopic embeddings alongside temporal signal classification to distinguish between strong, weak, and noisy trends. This signal taxonomy enables more fine-grained detection of topic trajectories and improves the forecasting of future relevance. Their temporal signal classification module has been directly adopted in our system.

To provide a broader perspective, \cite{huang2023survey} offer a comprehensive survey of AutoML methods and systems for clustering. They categorize existing techniques based on the components of the clustering pipeline being automated, such as data preprocessing, algorithm selection, hyperparameter optimization, and validation. The survey underscores the challenges inherent in AutoML for clustering, including the absence of universal evaluation metrics and the difficulty in benchmarking across diverse objectives and data types, thus identifying key areas for future research.

Efficiency and scalability are key considerations in the AutoML4Clust framework proposed by \cite{zhao2022automl4clust}. This system optimizes the entire clustering pipeline using efficient search strategies and integrates internal validation metrics to guide the search in the absence of ground truth labels. Its demonstrated performance on large-scale and high-dimensional datasets highlights its practicality for real-world unsupervised learning scenarios.

The effectiveness of meta-learning in AutoML for clustering heavily relies on the quality of data characterization measures. Authors of \cite{bengio2022evaluating} investigate a range of meta-features and assess their ability to predict the most suitable clustering algorithms. Their findings underscore the importance of selecting informative meta-features to improve the accuracy of meta-learners.

More recent work such as \cite{liu2021clustering} propose a meta-learning framework specifically designed for clustering algorithm recommendation. Their system extracts dataset meta-features to train a meta-learner that predicts clustering performance, enabling data-driven recommendations and reducing the need for manual experimentation.

Additional contributions \cite{lacoste2019clustering} presented a framework that automatically selects the best clustering algorithm for a given dataset based on a knowledge base populated through meta-learning. This approach is highly relevant to our AutoCluster methodology, where clustering models are selected based on prior knowledge and optimized using hyperparameter tuning.

Akiba et al. \cite{akiba2019optuna} introduced Optuna, a flexible and efficient hyperparameter optimization framework featuring a define-by-run interface, pruning strategies, and support for optimization algorithms such as Tree-structured Parzen Estimator (TPE) and Covariance Matrix Adaptation Evolution Strategy (CMA-ES). Widely used in AutoML and deep learning, Optuna is the optimization backend in our system.

The literature reveals key limitations, including the absence of automated topic modeling and labeling methods. Existing approaches often overlook the integration of hyperparameter optimization in clustering and topic modeling. Furthermore, there is a notable lack of automation in time series analysis. These gaps highlight the need for end-to-end AutoML pipelines in trend prediction tasks

\section{Methodology}

\begin{figure*}[h!]
    \centering
    \includegraphics[width=\textwidth, height=0.2\textheight, keepaspectratio]{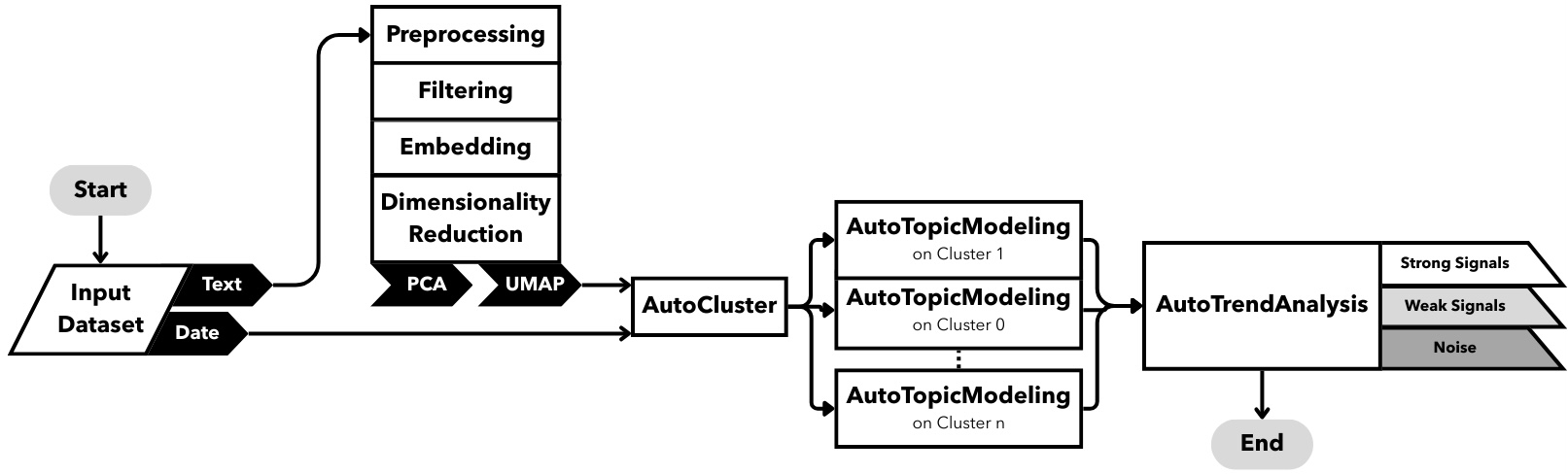}
    \caption{Methodology Diagram}
    \label{fig:methodology}
\end{figure*}

This section presents an extensive overview of the framework stages and flow. Figure \ref{fig:methodology} displays the framework Methodology Diagram.

\subsection{Input Dataset}

To ensure broad applicability, the input dataset requires only two fields: text and date. Texts should be topically similar to preserve the coherence of embeddings and topics—mixing unrelated sources (e.g., movie scripts and research papers) is discouraged. The date field can vary in granularity (e.g., year or year-month), with the final time resolution determined during the AutoTrendAnalysis phase. This design supports flexibility while maintaining trend analysis quality.

\subsection{Preprocessing and Filtering}

To prepare the text data for modeling, named entities (e.g., people, locations, organizations) that do not contribute to semantic trend analysis were removed. Text is then normalized by lowercasing, removing punctuation, short tokens, digits, and stop words. Lemmatization reduces words to their base forms to unify word variants. Finally, extremely frequent (in \textgreater{}90\% of documents) or rare (appearing \textless{}2 times) tokens were filtered out to reduce noise. This results in a clean, lemmatized token set optimized for embedding and topic modeling.

\subsection{Embedding and Dimensionality Reduction}

To capture semantic information from text, the pre-trained \texttt{allenai-SPECTER} model was employed, which is well-suited for scientific documents. SPECTER (Scientific Paper Embedding using Citation-informed TransformERs) was selected as our embedding model due to the scientific nature of the datasets employed in our experimentation. While the choice of embedding model is contingent upon the text source of the input dataset, the subsequent steps in our approach remain invariant to this initial selection. It generates high-dimensional sentence embeddings where semantically similar texts are positioned closely in the vector space. Since working directly with high-dimensional data can be computationally expensive and difficult to visualize, a two-step dimensionality reduction is applied. First, Principal Component Analysis (PCA) is used to reduce noise and preserve the most informative components. Then, Uniform Manifold Approximation and Projection (UMAP) projects the embeddings into two dimensions, maintaining both local and global structure. These low-dimensional vectors are then appended to the original dataset for clustering and visualization. This pipeline ensures both efficient computation and meaningful representation of textual data.

\subsection{AutoCluster}
\begin{figure}[h!]
    \centering
    \includegraphics[width=0.5\textwidth]{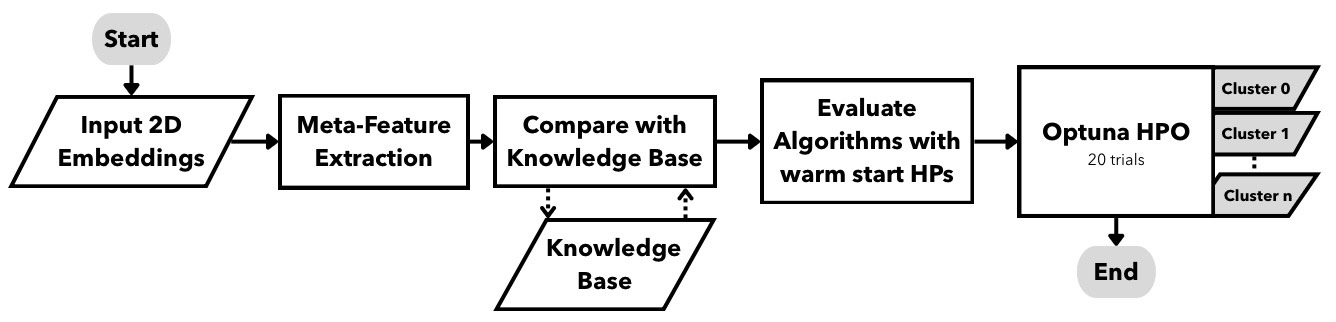}
    \caption{AutoCluster Diagram}
    \label{fig:autocluster}
\end{figure}

To generate high-quality clusters for topic modeling, our approach relies on 2D reduced embeddings and dataset meta-features that are independent of the dataset's dimensions. As shown in Figure \ref{fig:autocluster}, AutoCluster begins by extracting six meta-features from the 2D input space and matching them against a knowledge base built from 1,000 synthetic datasets. Each synthetic dataset consists of 100×2 matrices populated with random numerical values. Nine clustering algorithms were applied to each and optimized them using Optuna (10 trials) based on silhouette score. Valid configurations were retained only if they met three criteria: silhouette \textgreater{} 0.3, Davies-Bouldin index \textless{} 1.0, and a cluster count between 3 and 15. Each validated entry stores the optimal algorithm, meta-features, and hyperparameters.

During inference, the system retrieves the top three knowledge base matches via cosine similarity and evaluates each suggested algorithm (with warm-started parameters) on the target dataset. The best-performing algorithm is then passed to the final optimization phase.

20 Optuna trials were conducted to fine-tune hyperparameters, guided by a penalty-based constraint that discourages invalid cluster counts (\textless{}4 or \textgreater{}15). The trial maximizing the adjusted silhouette score is selected, producing robust and interpretable clusters for downstream topic modeling.

\subsection{AutoTopicModeling}
Effective topic modeling over large clustered datasets is computationally intensive, especially when comparing multiple algorithms with differing hyperparameter sensitivities and time complexities. To address this, our approach implements a successive halving strategy that reduces computational load by progressively eliminating underperforming models based on coherence scores.

\begin{figure}[h!]
    \centering
    \includegraphics[width=0.5\textwidth, height=0.15\textheight, keepaspectratio]{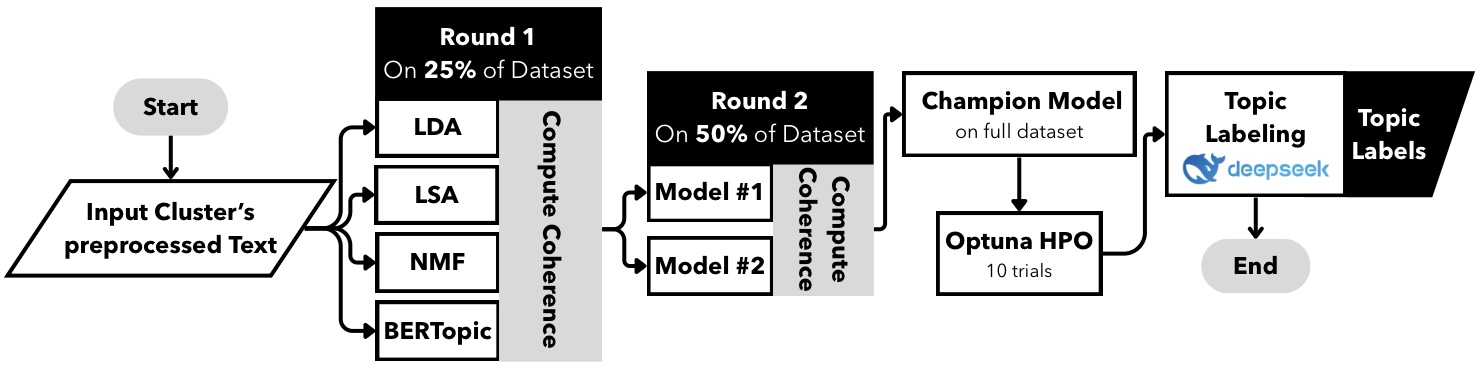}
    \caption{AutoTopicModeling Diagram}
    \label{fig:tm}
\end{figure}

As shown in Figure \ref{fig:tm}, our AutoTopicModeling pipeline proceeds in three stages: (1) Algorithm Selection, (2) Hyperparameter Optimization, and (3) Semantic Labeling. In the first stage, four candidate models—LDA, NMF, LSA, and BERTopic—are evaluated using a 25\% subset of the cluster. The two best performers (based on coherence score) are promoted to a second round using 50\% of the data. A penalty-adjusted coherence metric is applied to normalize score inflation from longer topic word lists: 5\% is deducted per word beyond 10 (up to 50\%).

The top model is then passed to Optuna for Bayesian hyperparameter optimization over 10 trials, maximizing coherence score. The final output includes the top 8 topic words per cluster, appended to the original dataset under a new \texttt{topic} column.

To generate interpretable labels, we apply the DeepSeek API to each list of topic words, using the following prompt: \texttt{"Analyze these topic words generated from search subject: '{subject}'.
For each topic, provide ONE concise label (2 to 6 words)"}, where \texttt{'{subject}'} is the subject of the dataset.

This stage incorporates Large Language Model (LLM)-based capabilities including: (1) Semantic Merging of near-duplicate topics; (2) Contextual Understanding informed by the subject domain; (3) Label Consistency via 5–30 total topics; and (4) Domain-Aware Relatedness for disambiguating polysemous terms.

Final labels are appended to the original data in a new column \texttt{topic\_label}. This end-to-end pipeline ensures scalable, efficient, and interpretable topic modeling tailored to each cluster, balancing algorithmic rigor with semantic clarity.

\subsection{AutoTrendAnalysis}

\begin{figure*}[h!]
    \centering
    \includegraphics[width=\textwidth]{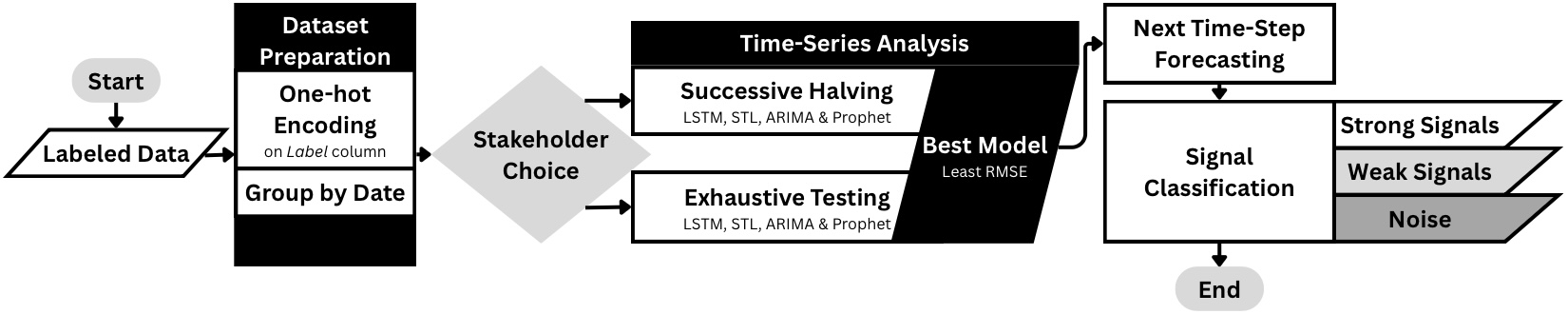}
    \caption{AutoTrendAnalysis Diagram}
    \label{fig:ta}
\end{figure*}

The final stage of the pipeline, \textbf{AutoTrendAnalysis}, integrates temporal dynamics into topic evolution by forecasting future trends using both classical and neural time-series models. Building on clustered and labeled data, it predicts topic prominence over time, providing interpretable insights (Figure~\ref{fig:ta}).

\subsubsection{Data Preparation}

Topic labels are one-hot encoded into binary indicator columns, where each document row flags the presence of topics. These indicators are aggregated by \texttt{date}, summing document counts to produce topic frequency time series. The resulting dataset retains only the \texttt{date} and topic columns, sorted and indexed by date. This format supports models such as ARIMA, STL, Prophet, and LSTM.

\subsubsection{Model Selection Strategies}

Two approaches are supported:

\begin{itemize}
    \item \textbf{Exhaustive Evaluation (default)}: All four models are trained on the first 80\% of data and evaluated on the final 20\%, using RMSE as the metric.
    \item \textbf{Successive Halving}: Initial evaluation on the latest 25\% identifies the top two models, which are then re-evaluated on 50\%. The best model by RMSE is selected and retrained on the full dataset in an 80-20 train-test split.
\end{itemize}

\subsubsection{Next Time-Step Forecasting}

The selected model forecasts the next time-step’s topic frequencies, configurable as weekly, monthly, or yearly, enabling adaptive temporal resolution per application.

\subsubsection{Signal Classification}

Forecasted topic values are classified into three groups:

\begin{itemize}
    \item \textbf{Strong Signals}: Values above the 50th percentile, indicating likely sustained or emerging prominence.
    \item \textbf{Weak Signals}: Values between the 10th and 50th percentile with positive trends, representing potential early indicators.
    \item \textbf{Noise}: Values below the 10th percentile or non-increasing trends, suggesting limited future relevance.
\end{itemize}

This classification highlights topics by momentum and relevance, supporting informed decision-making. The pipeline balances statistical rigor with semantic interpretability, producing actionable trend forecasts for downstream use.

\section{Experimentation and Results}
\label{chap:result}

\subsection{Datasets}
Experiments used three datasets:

\begin{itemize}
\item \textbf{arXiv CS Papers (50,000):} Recent computer science papers filtered by \texttt{category\_code} starting with \texttt{"cs"} from the kaggle "arXiv Scientific Research Papers Dataset", including exact publication dates. 
\item \textbf{NIPS Papers (7,000):} Machine learning conference papers with only publication year available. Source: Kaggle "NIPS Papers" dataset.

\item \textbf{Software Engineering Papers (30,000):} Papers from multiple academic sources filtered by keywords and quality, with only publication year available.
\end{itemize}

Only abstracts/summaries and publication dates/years were used. Dataset size variation highlights the method’s modularity and adaptability, supporting evaluation of clustering and topic modeling techniques.

\subsection{Results}
We present results from three experiments conducted on three datasets. Experiment 1 applies the AutoClustering module, Experiment 2 evaluates AutoTopicModeling, and Experiment 3 reports on the AutoTrendAnalysis module, which generates the final trend predictions. For brevity, Computer Science, Software Engineering, and Neural Information Processing Systems datasets are abbreviated as CS, SE, and NIPS, respectively.

\subsubsection{Experiment 1: AutoClustering Results}

Agglomerative clustering was the top-performing algorithm selected by our meta-learning approach across all datasets, likely due to favorable alignment with the datasets’ characteristics. Table~\ref{tab:clustering_results} summarizes the results. These results indicate reasonable cluster separation, with CS showing the highest silhouette score and SE the lowest, though still acceptable for downstream tasks.

\begin{table}[ht]
%\scriptsize
\caption{Clustering Results for the Datasets}
\centering
\begin{tabular}{|p{1.75cm}|p{1.75cm}|p{1.75cm}|p{1.75cm}|}
\hline
\textbf{Metric} & \textbf{CS} & \textbf{SE} & \textbf{NIPS} \\
\hline
\textbf{Algorithm} & Agglomerative & Agglomerative & Agglomerative \\
\hline
\textbf{Optimized Hyperparameters} & n\_clusters = 4, linkage = average, affinity = l2 & n\_clusters = 8, linkage = average, affinity = l1 & n\_clusters = 8, linkage = ward \\
\hline
\textbf{Silhouette Score} & 0.4327 & 0.3784 & 0.4126 \\
\hline
\textbf{Davies-Bouldin Score} & 0.9250 & 0.7481 & 0.7606 \\
\hline
\textbf{Number of Clusters} & 4 & 8 & 8 \\
\hline
\end{tabular}
\label{tab:clustering_results}
\end{table}

\subsection{Experiment 2: AutoTopicModeling Results}

The second experiment evaluates the performance of the champion topic modeling models on clusters obtained from the AutoClustering step for the three datasets. Table \ref{tab:autotopicmodeling_results} shows the obtained champion model per cluster for each dataset. 

\begin{table}[ht]
%\scriptsize
\caption{Champion Topic Modeling Algorithms per Cluster}
\centering
\renewcommand{\arraystretch}{1.3}
\begin{tabular}{|c|c|c|c|}
\hline
\textbf{Cluster} & \textbf{CS} & \textbf{SE} & \textbf{NIPS} \\
\hline
0 & NMF & LDA & NMF \\
1 & NMF & BERTopic & BERTopic \\
2 & BERTopic & NMF & BERTopic \\
3 & LDA & NMF & NMF \\
4 & - & NMF & NMF \\
5 & - & LDA & NMF \\
6 & - & LDA & NMF \\
7 & - & BERTopic & NMF \\
\hline
\end{tabular}
\label{tab:autotopicmodeling_results}
\end{table}

These results highlight that model effectiveness varies not only across datasets but also between clusters within the same dataset, reinforcing the importance of per-cluster model selection in topic modeling pipelines. Topic labels are displayed in Tables \ref{tab:cs}, \ref{tab:se} and \ref{tab:nips}.

\subsection{Experiment 3: AutoTrendAnalysis Results}

In this experiment, we apply AutoTrendAnalysis to forecast the evolution of research topics in three datasets. The goal is to predict the number of documents published per topic in future time steps. The experiment has two setups: first, applying AutoTrendAnalysis only to the CS dataset at weekly, monthly, and yearly resolutions, leveraging its fine-grained dates; second, applying it to all datasets using a uniform yearly time-step. In both, topics are classified as strong, weak, or noise signals based on forecasted number of documents.

\vspace{20pt}
\begin{table}[H]
%\scriptsize
\caption{Forecasted Trends for CS Dataset across Weekly, Monthly, and Yearly Time-Steps with Signal Classification}
\centering
\renewcommand{\arraystretch}{1.1}
\setlength{\tabcolsep}{4pt}
\begin{tabular}{|l|c|c|c|c|c|c|}
\hline
\textbf{\makecell{Topic}} & 
\multicolumn{2}{c|}{\textbf{1 Week}} & 
\multicolumn{2}{c|}{\textbf{1 Month}} & 
\multicolumn{2}{c|}{\textbf{1 Year}} \\
\hline
\textbf{\makecell{Forecast Date}} & \multicolumn{2}{c|}{06/02/2025} & \multicolumn{2}{c|}{01/02/2025} & \multicolumn{2}{c|}{01/01/2026} \\
\hline
 & \textbf{Docs} & \textbf{Signal} & \textbf{Docs} & \textbf{Signal} & \textbf{Docs} & \textbf{Signal} \\
\hline
\makecell[l]{Large Language Models} & 8.31 & Strong & 86.91 & Strong & 2604.31 & Strong \\
\makecell[l]{Text Classification} & 6.66 & Strong & 86.45 & Strong & 1311.42 & Strong \\
\makecell[l]{Learning Methods} & 4.23 & Strong & 56.91 & Strong & 903.95 & Strong \\
\makecell[l]{Human-Machine Learning} & 3.91 & Strong & 63.95 & Strong & 948.66 & Strong \\
\makecell[l]{Machine Learning Models} & 3.52 & Strong & 57.55 & Strong & 1015.22 & Strong \\
\makecell[l]{Traffic Prediction} & 2.30 & Strong & 49.93 & Strong & 682.29 & Strong \\
\makecell[l]{Optimization Algorithms} & 2.07 & Strong & 46.53 & Strong & 719.22 & Strong \\
\makecell[l]{Graph Learning} & 1.83 & Strong & 54.70 & Strong & 856.40 & Strong \\
Anomaly Detection & 1.04 & Strong & 7.37 & Noise & 122.44 & Noise \\
Federated Learning & 0.98 & Strong & 11.39 & Noise & 183.24 & Weak \\
\makecell[l]{Deep Neural Networks} & 0.97 & Strong & 50.67 & Strong & 644.98 & Strong \\
Reinforcement Learning & 0.97 & Strong & 27.68 & Strong & 439.45 & Strong \\
Knowledge Graphs & 0.93 & Weak & 26.45 & Weak & 357.31 & Weak \\
\makecell[l]{Neural Network\\Architecture} & 0.76 & Weak & 49.97 & Strong & 777.11 & Strong \\
Machine Translation & 0.15 & Weak & 19.32 & Weak & 284.26 & Weak \\
Object Detection & 0.04 & Weak & 21.29 & Weak & 309.87 & Weak \\
Scene Estimation & 0.04 & Weak & 11.65 & Weak & 152.68 & Noise \\
Cross-Domain Learning & 0.03 & Weak & 29.29 & Strong & 513.63 & Strong \\
Attention-Based Learning & -0.00 & Noise & 7.58 & Noise & 229.45 & Weak \\
Image Processing & -0.01 & Noise & 19.20 & Weak & 271.81 & Weak \\
Reinforcement Policy & -0.01 & Noise & 22.46 & Weak & 330.31 & Weak \\
Dialogue Generation & -0.01 & Noise & 11.36 & Noise & 198.26 & Weak \\
Bandit Algorithms & -0.04 & Noise & 11.39 & Weak & 146.36 & Noise \\
Time Series Forecasting & -0.10 & Noise & 16.54 & Noise & 329.19 & Weak \\
\hline
 \textbf{\makecell{RMSE}} & \multicolumn{2}{c|}{0.751} & \multicolumn{2}{c|}{1.175} & \multicolumn{2}{c|}{122.125} \\
\hline
\end{tabular}
\label{tab:cs}
\end{table}

\begin{table}[H]
%\scriptsize
\caption{Forecasted Trends for SE Dataset}
\centering
\renewcommand{\arraystretch}{1.0}
\begin{tabular}{|c|c|}
\hline
\multicolumn{2}{|c|}{\textbf{Time-Step: 1 Year, Date: 01-01-2024}} \\
\hline
\textbf{Topic} & \textbf{Forecasted Docs} \\
\hline
\multicolumn{2}{|c|}{\textbf{Strong Signals}} \\
\hline
Deep Learning Models & 202.56 \\
Open Source Ecosystem & 176.26 \\
Software Development & 173.01 \\
Software Safety Engineering & 151.02 \\
Business Process Modeling & 137.92 \\
Data Modeling Methods & 128.89 \\
Big Data Analytics & 106.20 \\
Mobile App Analysis & 98.19 \\
Security Vulnerabilities & 95.40 \\
Edge Computing & 83.29 \\
Model Verification & 80.74 \\
Defect Prediction Model & 76.06 \\
Autonomous Testing & 70.23 \\
Water Management Models & 62.97 \\
Algorithm Synthesis & 59.19 \\
Material Simulation & 57.11 \\
Fault Localization & 56.74 \\
\hline
\multicolumn{2}{|c|}{\textbf{Weak Signals}} \\
\hline
Mutation Testing & 46.23 \\
Static Code Analysis & 44.06 \\
Programming Languages & 43.73 \\
Constraint Solving & 39.48 \\
Formal Model Checking & 38.34 \\
Code Smells & 35.95 \\
Process Simulation & 33.60 \\
Algorithm Complexity & 31.94 \\
Concurrency Bugs & 30.82 \\
Program Verification & 20.44 \\
Network Algorithms & 20.14 \\
Logic And Proof & 20.07 \\
Symbolic Execution & 19.37 \\
Structural Optimization & 18.58 \\
\hline
\multicolumn{2}{|c|}{\textbf{Noise Signals}} \\
\hline
Finite Element Method & 15.68 \\
Quantum Computing & 13.82 \\
Graph Processing & 7.37 \\
Type Systems & -0.04 \\
\hline
\multicolumn{2}{|c|}{\textbf{RMSE:} 17.49} \\
\hline
\end{tabular}
\label{tab:se}
\end{table}

\begin{table}[H]
\scriptsize
\caption{Forecasted Trends for NIPS Dataset}
\centering
\renewcommand{\arraystretch}{1.0}
\begin{tabular}{|c|c|}
\hline
\multicolumn{2}{|c|}{\textbf{Time-Step: 1 Year, Date: 01-01-2018}} \\
\hline
\textbf{Topic} & \textbf{Forecasted Docs} \\
\hline
\multicolumn{2}{|c|}{\textbf{Strong Signals}} \\
\hline
Inference Methods & 69.00 \\
Sparse Estimation & 48.95 \\
Matrix Completion & 27.51 \\
Gradient Optimization & 19.79 \\
\hline
\multicolumn{2}{|c|}{\textbf{Noise Signals}} \\
\hline
Feature Selection & 14.02 \\
Generative Models & 12.81 \\
Graph Clustering & 12.31 \\
Online Optimization & 8.44 \\
\hline
\multicolumn{2}{|c|}{\textbf{RMSE:} 7.099} \\
\hline
\end{tabular}
\label{tab:nips}
\end{table}

\subsection{Discussion}
\label{sec:discussion}

Forecasted trends across datasets and time scales reveal distinct patterns in emerging and declining research topics.

\paragraph{CS Dataset (Weekly to Yearly).}
Table~\ref{tab:cs} shows LLMs as consistently strong signals, growing from 8.31 weekly to 2604.31 yearly documents, reflecting sustained interest in generative language models. Other topics like \textit{Text Classification}, \textit{Human-Machine Learning}, and \textit{Graph Learning} also maintain strong signals across scales. Notably, \textit{Neural Network Architecture} shifts from weak weekly to strong yearly signals, highlighting the benefit of long-term trend aggregation.

\paragraph{SE Dataset.}
Table~\ref{tab:se} shows strong signals in \textit{Deep Learning Models}, \textit{Open Source Ecosystem}, and \textit{Software Development}, indicating growing AI integration in software engineering. Emerging weak signals such as \textit{Mutation Testing} and \textit{Static Code Analysis} suggest potential shifts in quality assurance.

\paragraph{NIPS Dataset.}
As per Table~\ref{tab:nips}, \textit{Inference Methods} and \textit{Sparse Estimation} are strong signals, emphasizing foundational probabilistic modeling research. Topics like \textit{Feature Selection} and \textit{Graph Clustering} remain niche but relevant.

\subsection{Summary of Experimentation and Results}

We evaluated AutoClustering, AutoTopicModeling, and AutoTrendAnalysis on three datasets: CS, NIPS and SE papers, to uncover clusters, topics, and future research trends.

Experiment 1 showed that clustering quality varied, with CS yielding the clearest clusters, NIPS balanced results, and SE more challenging boundaries. Experiment 2 identified NMF as the preferred topic model in most CS and NIPS clusters, while SE exhibited more diverse algorithm preferences, including LDA. Experiment 3’s trend forecasts classified topics as strong, weak, or noise signals. Foundational AI topics, especially LLMs, showed sustained growth in CS, while SE emphasized applied areas like Deep Learning and Open Source. The NIPS dataset highlighted enduring theoretical interests in inference and sparse methods.

Overall, multi-scale forecasting reveals evolving research trajectories, offering valuable insights to guide future research, funding, and strategic planning.

\section{Conclusion, Limitations and Future Work}

This paper presents an automated, scalable, and interpretable AutoML framework for trend prediction from large-scale textual datasets. Leveraging meta-learning, optimized clustering, topic modeling, and neural time-series forecasting, the pipeline transforms raw temporal text data into actionable insights with minimal human intervention. Key contributions include AutoCluster for algorithm selection, successive halving with Optuna for topic modeling, and robust trend classification, enabling broad applicability and near real-time monitoring.

However, several limitations remain. Scalability may be challenged by extremely large or heterogeneous data without further optimization. The current system is limited to textual data and batch processing, restricting real-time adaptability and multimodal analysis. Topic labeling depends on existing APIs and coherence metrics, which may reduce interpretability for complex or overlapping topics. The fixed algorithm pool may miss emerging methods, and potential biases in training data could affect generalization.

Future work will address these gaps by enhancing scalability through distributed computing, integrating multimodal data sources, and enabling real-time trend updates. Incorporating external knowledge bases and exploring advanced forecasting models such as Transformer architectures with uncertainty quantification are also planned. Additionally, efforts will focus on improving fairness, transparency, and bias mitigation to ensure ethical deployment. These improvements aim to expand the framework’s accuracy, applicability, and ethical robustness in diverse real-world scenarios.

\bibliographystyle{IEEEtran}
\bibliography{references}
\end{document}